\documentclass[a4paper,11pt]{article}

\usepackage{jheppub} 

\usepackage[T1]{fontenc} 

\usepackage{amsfonts,amsmath} \usepackage{mathrsfs} \usepackage{bm} \usepackage{dcolumn}
\usepackage{epsfig} \usepackage{latexsym}
\usepackage{soul}
\usepackage[dvipsnames]{xcolor}
\usepackage{multirow}
\usepackage{setspace}

\newcommand\Ns{\mathcal{N}}
\newcommand\sgn{\text{sgn}}

\title{A route from maximal chaoticity to integrability}

\author[a,b]{Chen Ma}
\author[b,1]{and Chushun Tian \note{Corresponding author.}}

\affiliation[a]{School of Physical Sciences, University of Chinese Academy of Sciences, Beijing 100049, China}
\affiliation[b]{CAS Key Laboratory of Theoretical Physics and Institute of Theoretical Physics, \\Chinese Academy of Sciences, Beijing 100190, China}

\emailAdd{machen@itp.ac.cn}
\emailAdd{ct@mail.itp.ac.cn}

\abstract{
 We study the chaos exponent of some variants of the Sachdev-Ye-Kitaev (SYK) model, namely, the $\Ns=1$ supersymmetry (SUSY)-SYK model and its sibling, the $(N|M)$-SYK model which is not supersymmetric in general, for arbitrary interaction strength. We find that for large $q$ the chaos exponent of these variants, as well as the SYK and the $\Ns=2$ SUSY-SYK model, all follow a single-parameter scaling law. By quantitative arguments we further make a conjecture, i.e. that the found scaling law might hold for general one-dimensional (1D) SYK-like models with large $q$. This points out a universal route from maximal chaos towards completely regular or integrable motion in the SYK model and its 1D variants.}

\begin{document}
\maketitle
\flushbottom

\section{Introduction}
\label{sec:1}

The SYK model \cite{SY1993,Kitaev15}, that describes the quantum motion of $N$ Majorana fermions subjected to a random $q$-body mutual interaction, has attracted a lot of attentions in recent years. One of its most interesting features is that, in the strong coupling limit, it is exactly solvable and satisfies the so-called maximal chaos bound \cite{Maldacena16,Maldacena16b,Kitaev15}, i.e. the chaos exponent $\lambda_L$ is $\frac{2\pi}{\beta}$, where $\beta$ is the inverse temperature.\footnote{Throughout this work the Planck constant is set to unity.} Many variants of the SYK model have been introduced. Among the 1D variants is the SYK model with SUSY \cite{Peng17,Li17,Yoon17,Peng17b,Fu17,HJ18,HJ18b,Narayan18,Garcia18,Kato18,Sun20,Gates21,Gates22,Ahn22,Heydeman22,He22,Peng20}. It has been found that in 1D the $\Ns=1,2$ SUSY-SYK model \cite{Fu17,Peng17b} are also maximally chaotic. Another interesting 1D variant that is not supersymmetric in general, but closely related to SUSY-SYK models was introduced in \cite{Marcus18}. It consists of $N$ fermions and $M$ bosons --- thus dubbed the $(N|M)$-SYK model here --- and was found to be maximally chaotic in the strong coupling limit.

Less attentions have been paid to the SYK model and its variants beyond the strong coupling regime, where the conformal symmetry is strongly broken and $\lambda_L$ deviates substantially from the maximal chaos bound. In view of rich universal properties carried by the SYK(-like) models in the strong coupling limit, an interesting question is whether those models may exhibit certain non-maximal, but universal chaotic phenomena. In this work, we address this issue, with a focus on large $q$ \cite{Maldacena16,Fu17,Peng20,Bha17,Tar18,Jiang19,Choi19,Khram21,Len21,Bha22}.

To the best of our knowledge, such a task was first undertaken in \cite{Maldacena16}. There, the chaos exponent of the SYK model, which is composed merely of fermions (thus we shall not distinguish the terms of SYK model and $\Ns=0$ SUSY-SYK model below), with arbitrary interaction strength was calculated for large $q$. It was found to display single-parameter scaling behaviors. Specifically, when $\lambda_L$ is rescaled by the maximal chaos bound $2\pi/\beta$, one finds
\begin{align}
  \label{eq:0.1}
  \frac{\lambda_L}{2\pi / \beta} = v(\beta \mathcal J),
\end{align}
where the scaling function, $v(x)$, satisfies
\begin{align}
  \label{eq:0.2}
  \quad x = \frac{\pi v(x)}{\cos\left(\frac{\pi v(x)}{2}\right)}.
\end{align}
According to Eqs.(\ref{eq:0.1}) and (\ref{eq:0.2}), the model's detailed constructions, which are governed by the interaction strength $J$ and $q$, enter only into $\mathcal J$, i.e. the (dimensionless) scaling factor $\beta\mathcal J$. Interestingly, Eqs.(\ref{eq:0.1}) and (\ref{eq:0.2}) give the maximal chaos bound in the strong coupling limit $\beta\mathcal J \rightarrow \infty$, while give a vanishing $\lambda_L$ in the weak coupling limit $\beta\mathcal J \rightarrow 0$. In \cite{Peng20} it was found that Eqs.(\ref{eq:0.1}) and (\ref{eq:0.2}) hold also for the $\Ns=2$ SUSY-SYK model for large $q$, and the only difference lies in the explicit expression of $\mathcal J$ in terms of $J$ and $q$. So, a natural question is to what extent Eqs.(\ref{eq:0.1}) and (\ref{eq:0.2}) are universal.

In this work, we calculate the chaos exponent of the $\Ns=1$ SUSY-SYK and the $(N|M)$-SYK model with arbitrary $J$, but for large $q$. We find that for these two models Eqs.(\ref{eq:0.1}) and (\ref{eq:0.2}) still hold, and similar to the situations in the $\Ns=2$ SUSY-SYK model, the only difference is the dependence of $\mathcal J$ on $J$, $q$. Furthermore, by refining the derivations for the $\Ns=0,1,2$ SUSY- and $(N|M)$-SYK model, we conjecture that the single-parameter scaling law described by Eqs.(\ref{eq:0.1}) and (\ref{eq:0.2}) would hold for more general 1D variants of the SYK model. Thus Eqs.(\ref{eq:0.1}) and (\ref{eq:0.2}) encompass a universal route for going beyond maximal chaoticity and, in particular, towards integrability.

The remainder of the paper is organized as follows. In the next section we will calculate the so-called double commutator \cite{Witten17} for the $\Ns=1,2$ SUSY-SYK model. This commutator carries essentially the same information on chaos as the out-of-time-order correlator \cite{Kitaev15}, and is a straightforward generalization of the squared commutator $[p(t),p(0)]^2$ ($p$ being the Heisenberg momentum operator), which reduces to the stability matrix of the referenced phase-space trajectory in the semiclassical limit. The latter commutator was introduced by Larkin and Ovchinnikov and allowed them to discover quantum chaos in disordered superconductors \cite{Larkin1969}. We find the scaling law Eqs.(\ref{eq:0.1}) and (\ref{eq:0.2}) for the $\Ns=1$ SUSY-SYK model, and reproduce a result \cite{Peng20} for the $\Ns=2$ SUSY-SYK model. In Sec. \ref{sec:3} we introduce the $(N|M)$-SYK model and calculate a variety of Green functions of this model. With these preparations we study the non-maximal chaoticity of this model in Sec. \ref{sec:4}. In particular, we derive Eqs.(\ref{eq:0.1}) and (\ref{eq:0.2}). In Sec. \ref{sec:5} we make a conjecture about the validity of Eqs.(\ref{eq:0.1}) and (\ref{eq:0.2}) in more general 1D SYK-like models by refining the derivations in Secs. \ref{sec:2}-\ref{sec:4}. We conclude in Sec. \ref{sec:7}. Some technical details are shuffled to Appendix \ref{app:2} and \ref{app:1}.

\section{Non-maximal chaos in the \texorpdfstring{$\boldsymbol{\Ns=1,2}$}{N=1,2} SUSY-SYK model}
\label{sec:2}

In this section, we briefly introduce the $\Ns=1,2$ SUSY-SYK model and compute their double commutators. This allows us to calculate their chaos exponents for arbitrary interaction strength.

\subsection{\texorpdfstring{$\boldsymbol{\Ns=1}$}{N=1}}
\label{sec:2.1}

\subsubsection{Description of the model}
\label{sec:2.1.1}

The $\Ns=1$ SUSY-SYK model was introduced in \cite{Fu17}. It consists of $N$ Majorana fermions, each of which interacts with $(q-1)$ Majorana fermions\footnote{For the SUSY- and $(N|M)$-SYK model $q$ is assumed to be odd, while for the SYK model it is assumed to be even.}, with a coupling constant $C_{i_1\dots i_q}$ antisymmetric with respect to the indices $i_1,\dots,i_q$. These coupling constants are independent Gaussian random variables, all of which can be set to real. They have zero mean and variance
\begin{align}
  \label{eq:2.3}
  \overline{C_{i_1\dots i_q}^2} = \frac{(q-1)!J^2}{N^{q-1}},
\end{align}
where the overline stands for the disorder average. The Lagrangian is
\begin{align}
  \label{eq:2.4}
  \mathcal L = \int d\theta \Big(\Psi_i D\Psi_i + \frac{i^{\frac{q-1}{2}}}{q!} C_{i_1\dots i_q} \Psi_{i_1}\dots \Psi_{i_q}\Big).
\end{align}
Here the superfield
\begin{align}
  \label{eq:2.5}
  \Psi_i = \Psi_i(t,\theta) \equiv \psi_i(t) + \theta b_i(t)
\end{align}
represents the Majorana fermion $i$; it is defined over the supersymmetric spacetime, which has a time direction, $t$, and a Grassmann direction, $\theta$. The factor $i^{\frac{q-1}{2}}$ ensures the reality of the Lagrangian, and $D = \partial_\theta - i\theta\partial_t$ is the covariant derivative. Note that Einstein's summation convention is implied hereafter and $q!$ arises from the total antisymmetry of $C$. The first term is the kinetic term and the second accounts for the fermion interaction. According to Eq.(\ref{eq:2.4}) $C^2_{i_1\dots i_q}$ and thereby $J^2$ --- instead of $C_{i_1\dots i_q}$ and $J$ in the SYK model --- have the unit of energy.

\begin{figure}[hb]
  \centering
  \includegraphics[width=0.5\textwidth]{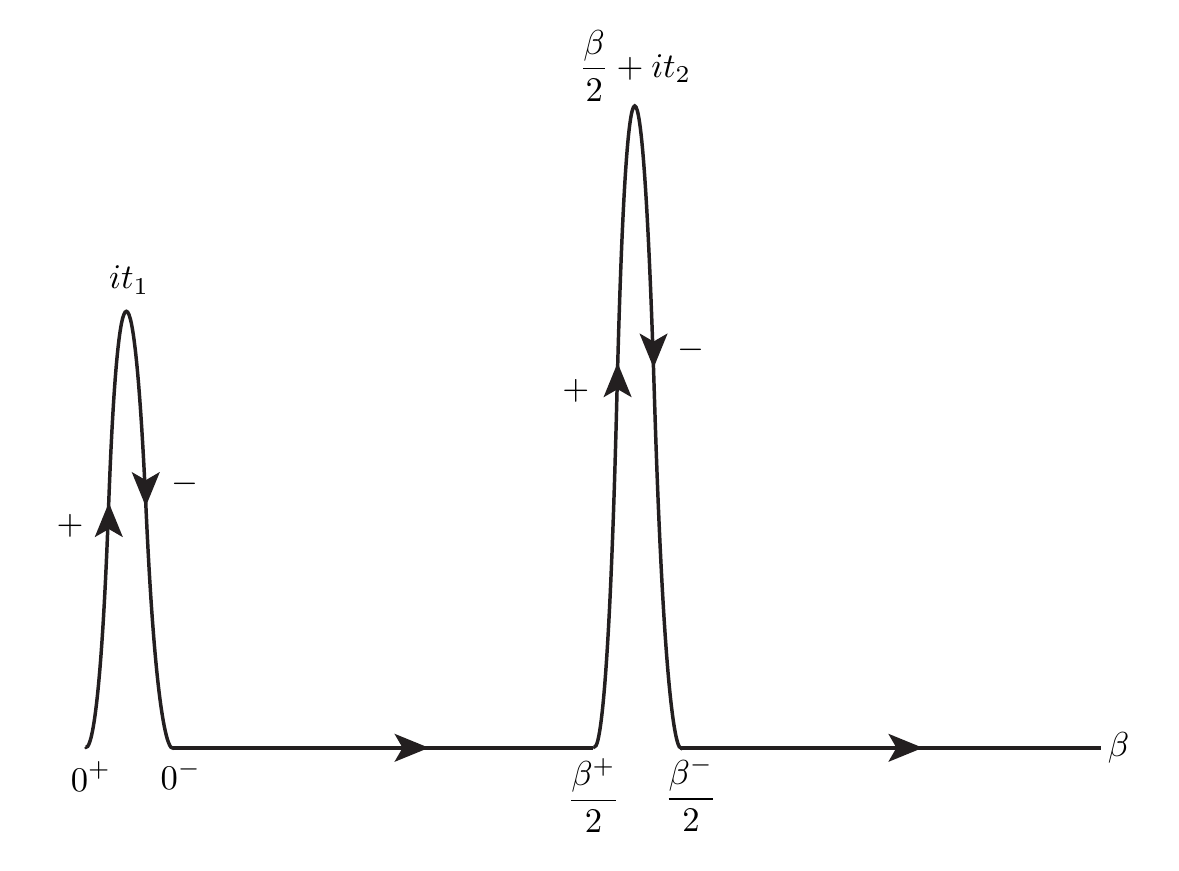}
  \caption{The contour $\mathcal C$ in the definition of the double commutator.}
  \label{fig:0}
\end{figure}

\subsubsection{The Bethe-Salpeter equation for the double commutator}
\label{sec:2.1.2}

To probe chaotic phenomena we introduce the double commutator defined as
\begin{align}
  \label{eq:2.6}
  &\frac{1}{N} F(\xi_1;\xi_2) \equiv \frac{1}{N^2} \overline{\langle \Phi_i(0,0)\Psi_j(it_1,\theta_1) \Phi_i(\frac{\beta}{2},0) \Psi_j(\frac{\beta}{2} + it_2,\theta_2) \rangle_{\mathcal C}}
\end{align}
with
\begin{align}
  \label{eq:2.6.1}
  \Phi_i(\tau,0) \equiv \Psi_i(\tau^+,0) - \Psi_i(\tau^-,0).
\end{align}
Here $\xi_i \equiv(t_i,\theta_i)$ with $t_i\in \mathbb R^+$ and $\theta_i$ being the Grassmann coordinate, and $\langle\dots\rangle_{\mathcal C}$ denotes the functional average on the contour $\mathcal C$ in the complex time plane as shown in Fig.\ref{fig:0}, i.e.
\begin{align}
  \label{eq:2.7}
  \langle\dots\rangle_{\mathcal C} \equiv \frac{\int\mathscr D [\{\Psi_i\}] e^{\int_\mathcal C dt \mathcal L} (\dots) }{\int\mathscr D [\{\Psi_i\}] e^{\int_\mathcal C dt \mathcal L}}.
\end{align}
The contour has two rails, each of which has two sides. The left (right) is denoted as $+$ ($-$). Along the contour the real part of the complex time keep increasing. As a result, the contour-ordered propagator can be obtained from the Euclidean propagator, $G(\xi_1|\xi_2)$, by analytic continuation.

\begin{figure}
  \centering
  \includegraphics[width=0.7\textwidth]{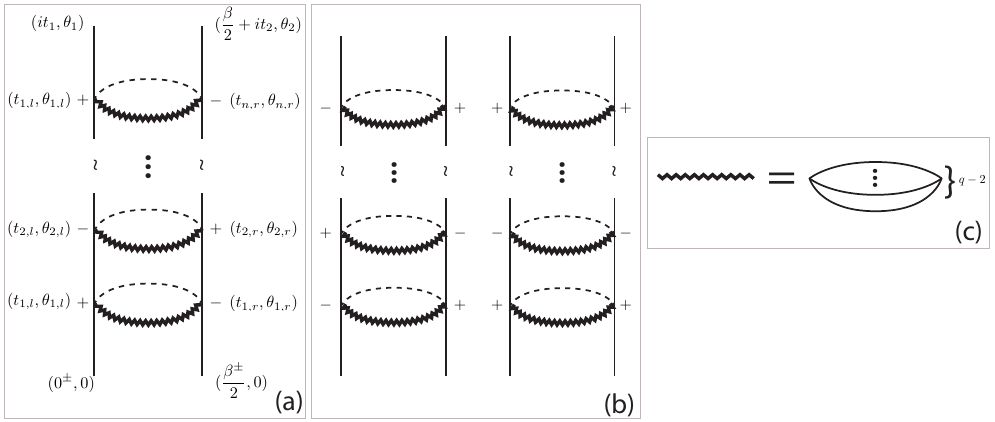}
  \caption{An example of the diagram for the double commutator of the $\Ns=1$ SUSY-SYK model (a). Two more examples are shown in (b), where the superspacetime coordinates of the external fields and interaction vertexes are omitted. In each diagram, there are two vertical rails and many rungs. Each rung is composed of $(q-2)$ Wightman propagators (c) and an interaction line.}
  \label{fig:6.1}
\end{figure}

Below we generalize the diagrammatic method developed in \cite{Witten17} to calculate Eq.(\ref{eq:2.6}). For $N\gg 1$ the double commutator $F$ is dominated by the diagrams exemplified in Fig. \ref{fig:6.1}(a) and (b). Each diagram has two vertical rails, with the top of the right (left) rail being the external field $\Psi_j(\frac{\beta}{2}+it_2,\theta_2)$ ($\Psi_j(it_1,\theta_1)$) and the bottom of the right (left) rail being $\Psi_i(\frac{\beta^\pm}{2},0)$ ($\Psi_i(0^\pm,0)$). In each rail, $n$ interaction vertexes are inserted, which are denoted as $a_1, a_2, \dots, a_n$ ($b_1, b_2, \dots, b_n$) from the bottom to the top in the right (left) rail. The superspacetime coordinate of $a_k$ ($b_k$) is denoted as $\xi_{k,r} = (t_{k,r},\theta_{k,r})$ ($\xi_{k,l}= (t_{k,l},\theta_{k,l})$). The vertexes are arranged in the way that all complex times $t_{k,r}$ ($t_{k,l}$) are located on the right (left) rail of the contour $\mathcal C$ and $\text{Im} t_{n,r} > \text{Im} t_{n-1,r} > \cdots >\text{Im} t_{1,r}$ ($\text{Im} t_{n,l} > \text{Im} t_{n-1,l} > \cdots >\text{Im} t_{1,l}$) on the right (left) rail. The vertex $a_k$ in the right rail is connected to the vertex $b_k$ in the left via a horizontal rung, which is made up of an interaction line (dashed line) and a zigzag line, representing [Fig. \ref{fig:6.1}(c)] the product of $(q-2)$ Wightman propagators $G_W(it_1,\theta_1|it_2,\theta_2) = G(it_1,\theta_1|it_2+\frac{\beta}{2},\theta_2)$ ($t_{1,2}\in\mathbb R$), each of which is obtained from the Euclidean propagator $G$ (thin solid line) by analytic continuation.

Because each rail in the contour $\mathcal C$ has two sides, when a vertex is attached to a rail, its complex time can be located on either the $+$ or $-$ side of the rail (see Fig. \ref{fig:6.1}(a) and (b) for examples). Importantly, since along the contour direction $dt$ increases (decreases) on the $+$ ($-$) side of a rail, contributions from two diagrams differing only on which side of a vertex is inserted into a specific rail differ only in the sign. Moreover, owing to the structures of $\Phi_i(0,0)$ and $\Phi_i(\frac{\beta}{2},0)$ in the definition Eq.(\ref{eq:2.6}), contributions from the diagrams, such that the external fields located at the bottom of the left (right) rail contracting with the vertexes in the right (left) rail, must cancel out.

\begin{figure*}[tb]
  \centering
  \includegraphics[width=0.7\textwidth]{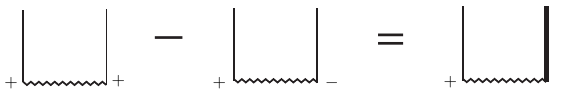}
  \caption{Diagrammatic illustration of the appearance of the retarded Green function (heavy solid line) in Eq.(\ref{eq:2.8.1}).}
  \label{fig:1.5}
\end{figure*}

\begin{figure*}[tb]
  \centering
  \includegraphics[width=0.6\textwidth]{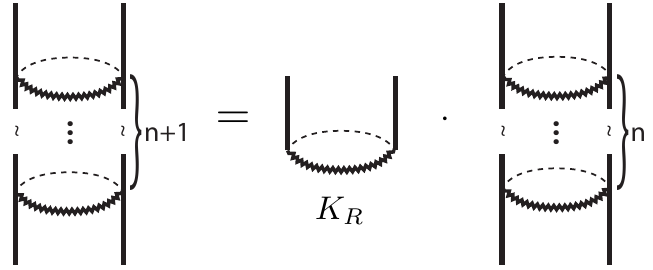}
  \caption{$(n+1)$-rung diagrams are generated by the acting of the retarded kernel $K_R$ on $n$-rung diagrams.}
  \label{fig:6.2}
\end{figure*}

Suppose that we would like to add one more rung connecting $a_{n+1}$ and $b_{n+1}$ to an $n$-rung diagram of $n$ rungs. We first fix $b_{n+1}$ in any side of the left rail. Because $a_{n+1}$ can be attached to any side of the right rail of the contour $\mathcal C$ and the Wightman propagator does not depend on to which side $a_{n+1}$ is attached, we obtain a contribution (Fig. \ref{fig:1.5})
\begin{align}
  \label{eq:2.8.1}
  &J^2(q-1)\int d{\xi_{n+1,r}} (G_W(\xi_{n+1,l};\xi_{n+1,r}))^{q-2} (-i\Theta(t_2 - \text{Im} t_{n+1,r} + i\theta_2\theta_{n+1,r}))\nonumber\\
  &\times(G(it_2,\theta_2|\frac{\beta^-}{2} + i\text{Im} t_{n+1,r},\theta_{n+1,r}) - G(it_2,\theta_2|\frac{\beta^+}{2} + i\text{Im} t_{n+1,r},\theta_{n+1,r}))\nonumber\\
  &= J^2(q-1) \int d\xi_{n+1,r}(G_W(\xi_{n+1,l};\xi_{n+1,r}))^{q-2} G_R(\xi_2|\xi_{n+1,r}),
\end{align}
with $\Theta$ being the Heaviside function. Here the second line accounts for adding together two contributions obtained from attaching $a_{n+1}$ to the $+$ and $-$ side, respectively, that (when multiplied by the $\Theta$ function) gives rise to the retarded Green function $G_R$ (heavy solid line). Let us further allow the vertex $b_{n+1}$ to change from one side of the left rail to the other, and its superspacetime coordinate to vary. Be the same token we can promote the above contribution to
\begin{align}
  \label{eq:2.9.1}
  J^2(q-1) \int d\xi_{n+1,l} d\xi_{n+1,r} (G_W(\xi_{n+1,l};\xi_{n+1,r}))^{q-2} G_R(\xi_2|\xi_{n+1,r}) G_R(\xi_1|\xi_{n+1,l})
\end{align}
Therefore, upon summing up $n$-rung diagrams with different $+/-$ configurations in each rail, a diagram, denoted as $F_n$, results, which is obtained by replacing the thin solid lines in the vertical rails of Fig. \ref{fig:6.1} by heavy solid lines (see Fig. \ref{fig:6.2}). Moreover, a diagram for $F_{n+1}$ can be generated from acting on a diagram for $F_n$ by the operation Eq.(\ref{eq:2.9.1}). More precisely, by introducing the retarded kernel
\begin{align}
  \label{eq:2.10.1}
  K_R(\xi_1;\xi_2|\xi_3;\xi_4) = J^2(q-1) G_R(\xi_1|\xi_3)G_R(\xi_2|\xi_4) (G_W(\xi_3;\xi_4))^{q-2}
\end{align}
defined via the integrand of Eq.(\ref{eq:2.9.1}), where we have made the change of variables in Eq.(\ref{eq:2.9.1}): $\xi_{n+1,l}\rightarrow \xi_3,\xi_{n+1,r}\rightarrow \xi_4$, (Note that this kernel holds for general $q$.) we have
\begin{align}
  \label{eq:6.3}
  F_{n+1}(\xi_1;\xi_2) = \int d\xi_4 d\xi_3 K_R(\xi_1;\xi_2|\xi_3;\xi_4) F_n(\xi_3;\xi_4)
\end{align}
with $d\xi_i \equiv dt_i d\theta_i$. Here the order of the measures: $d\xi_4$ and $d\xi_3$ is in accordance with the contour order. The diagrammatic illustration of Eq.(\ref{eq:6.3}) is given in Fig. \ref{fig:1.5}.

In order to show a relation between the $\Ns=1$ SUSY-SYK model and $(N|M)$-SYK model in Sec. \ref{sec:4}, we expand $F_n$, $G_R$ and $G_W$ in Grassmann variables as
\begin{align}
  \label{2.11.1}
  \left(\!\!
    \begin{array}{c}
      F_n(\xi_1;\xi_2) \\
      G_R(\xi_1|\xi_2) \\
      G_W(\xi_1;\xi_2) \\
    \end{array}
  \!\!\right)
   \!\equiv\! \left(\!\!
    \begin{array}{c}
      F_n^\psi(t_1;t_2) \\
      G_R^\psi(t_1|t_2) \\
      G_W^\psi(t_1;t_2) \\
    \end{array}
  \!\!\right)\!+\!\theta_1\theta_2 \left(\!\!
    \begin{array}{c}
      F_n^b(t_1;t_2) \\
      G_R^b(t_1|t_2) \\
      G_W^b(t_1;t_2) \\
    \end{array}
  \!\!\right)\,\,\,\,\,
\end{align}
and integrate out the Grassmann variables $\theta_3$ and $\theta_4$ in Eq.(\ref{eq:6.3}). As a result, we have
\begin{align}
  \label{eq:2.12.1}
  F^\psi_{n+1}(t_1;t_2) = \int dt_4 dt_3 \Big(K_R^{(11)}(t_1;t_2|t_3;t_4) F^\psi_n(t_3;t_4) + K_R^{(12)}(t_1;t_2|t_3;t_4) F^b_n(t_3;t_4)\Big)
\end{align}
with
\begin{align}
  \label{2.13.1}
  K_R^{(11)}(t_1;t_2|t_3;t_4) &= J^2(q-1)(q-2) G_R^\psi(t_1|t_3) G_R^\psi(t_2|t_4) G_W^b(t_3;t_4) (G_W^\psi(t_3;t_4))^{q-3},\\
  \label{2.14.1}
  K_R^{(12)}(t_1;t_2|t_3;t_4) &= J^2(q-1) G_R^\psi(t_1|t_3) G_R^\psi(t_2|t_4) (G_W^\psi(t_3;t_4))^{q-2},
\end{align}
and
\begin{align}
  \label{eq:2.15.1}
  F^b_{n+1}(t_1;t_2) = \int dt_4 dt_3 K_R^{(21)}(t_1;t_2|t_3;t_4) F^\psi_n(t_3;t_4),
\end{align}
with
\begin{align}
  \label{eq:2.16.1}
  K_R^{(21)}(t_1;t_2|t_3;t_4) = -J^2(q-1) G_R^b(t_1|t_3) G_R^b(t_2|t_4) (G_W^\psi(t_3;t_4))^{q-2}.
\end{align}

Let us return to Eq.(\ref{eq:6.3}). Combining it with $F = \sum_{n=0}^{\infty} F_n$, we obtain the Bethe-Salpeter equation
\begin{align}
  \label{eq:2.8}
  F(\xi_1;\xi_2) = F_0(\xi_1,\xi_2) + \int d\xi_4 d\xi_3 K_R(\xi_1;\xi_2|\xi_3;\xi_4) F(\xi_3;\xi_4)
\end{align}
for general $q$. For large $q$, the Wightman function is given by
\begin{align}
  \label{eq:2.C.16}
  G_W(\xi_1;\xi_2) = \frac{1}{2}\left(1+\frac{1}{q} \ln \frac{\pi v/(\beta \mathcal J)}{\cosh \frac{\pi v (t_1-t_2)}{\beta} + \frac{\pi v}{\beta} \theta_1\theta_2}\right),
\end{align}
where $v(\beta\mathcal J)$ is given by Eq.(\ref{eq:0.2}) and
\begin{align}
  \label{eq:2.13}
  \mathcal J = \frac{q J^2}{2^{q-2}}
\end{align}
is fixed, and the retarded Green function in Eq. (\ref{eq:2.10.1}) can be replaced by the free one, read
\begin{align}
  \label{eq:2.19.1}
  G_{R0}(\xi_1|\xi_2) = -i\Theta(t_1-t_2+i\theta_1\theta_2).
\end{align}
So Eq.(\ref{eq:2.10.1}) is simplified as
\begin{align}
  \label{eq:2.9}
  K_R(\xi_1;\xi_2|\xi_3;\xi_4) = G_{R0}(\xi_1|\xi_3) G_{R0}(\xi_2|\xi_4) W(\xi_3;\xi_4),
\end{align}
where
\begin{align}
  \label{eq:2.C.22}
  W(\xi_3;\xi_4) \equiv J^2(q-1) (G_W(\xi_3;\xi_4))^{q-2} = \frac{\pi v}{\beta} \frac{1}{\cosh\frac{\pi v(t_3-t_4)}{\beta} + \frac{\pi v}{\beta}\theta_3\theta_4}.
\end{align}

\subsubsection{Solution of the Bethe-Salpeter equation}
\label{sec:2.1.3}

Chaotic phenomena occur at large $t_1$ and $t_2$. So we can ignore the first term on the right-hand side of Eq.(\ref{eq:2.8}). By further performing the derivative: $D_1D_2$ on both sides, where $D_1$ acts on the argument $\xi_1$ and $D_2$ on $\xi_2$, we reduce Eq.(\ref{eq:2.8}) to
\begin{align}
  \label{eq:2.15}
  D_1D_2F(\xi_1;\xi_2) = W(\xi_1;\xi_2)F(\xi_1;\xi_2).
\end{align}
This reduces the Bethe-Salpeter equation to a differential equation, which we solve below.

We expand $W$ and $F$ in Grassmann variables $\theta_{1,2}$ as
\begin{eqnarray}
\label{eq:2.16}
  \left(\!\!
    \begin{array}{c}
      W(\xi_1;\xi_2) \\
      F(\xi_1;\xi_2) \\
    \end{array}
  \!\!\right)
   \!\equiv\! \left(\!\!
    \begin{array}{c}
      W^\psi(t_1;t_2) \\
      F^\psi(t_1;t_2) \\
    \end{array}
  \!\!\right)\!+\!\theta_1\theta_2 \left(\!\!
    \begin{array}{c}
      W^b(t_1;t_2) \\
      F^b(t_1;t_2) \\
    \end{array}
  \!\!\right).\,\,\,\,\,
\end{eqnarray}
With their substitution Eq.(\ref{eq:2.15}) reduces to
\begin{align}
  \label{eq:2.17}
  \big(\partial_{t_1}\partial_{t_2} - ((W^\psi)^{2}-W^b)\big) F^\psi= 0
\end{align}
and
\begin{align}
  \label{eq:2.18}
  F^b= -W^\psi F^\psi.
\end{align}
Substituting Eqs.(\ref{eq:2.C.22}) and (\ref{eq:2.18}) into Eq.(\ref{eq:2.17}), we obtain
\begin{align}
  \label{eq:6.45}
  \left[\partial_{t_1}\partial_{t_2} - \frac{2\pi^2v^2}{\beta^2} \frac{1}{\cosh^2 \frac{\pi v (t_1-t_2)}{\beta}}\right] F^\psi = 0.
\end{align}
In appendix \ref{app:2} we solve this equation. The solution is
\begin{align}
  \label{eq:2.20.1}
  F^\psi(t_1;t_2) = \frac{e^{\lambda_L \frac{t_1+t_2}{2}}}{\cosh\frac{\pi v(t_1-t_2)}{\beta}}.
\end{align}
Substituting it into Eqs.(\ref{eq:2.16})-(\ref{eq:2.18}), we obtain
\begin{align}
  \label{eq:2.20.2}
  F_b(t_1;t_2) = -\frac{\pi v}{\beta} \frac{e^{\lambda_L \frac{t_1+t_2}{2}}}{\cosh^2\frac{\pi v(t_1-t_2)}{\beta}}
\end{align}
and
\begin{align}
  \label{eq:2.21}
  F(\xi_1;\xi_2)=\frac{e^{\lambda_L \frac{t_1+t_2}{2}}}{\cosh \frac{\pi v (t_1-t_2)}{\beta}+\frac{\pi v}{\beta}\theta_1\theta_2}.
\end{align}

\subsection{\texorpdfstring{$\boldsymbol{\Ns=2}$}{N=2}}
\label{sec:2.2}

The out-of-time-order correlator of the $\Ns=2$ SUSY-SYK model has been studied in \cite{Peng20}. In this part we generalize the calculation scheme for the $\Ns=1$ SUSY-SYK model to calculate the double commutator in the case of $\Ns=2$. This would allow us to discover some similarities of this model's double commutator to that of the $\Ns=1$ SUSY-SYK model and the $(N|M)$-SYK model to be studied in the next two sections, and thus establish the universality of Eqs.(\ref{eq:0.1}) and (\ref{eq:0.2}) for more general models (see Sec. \ref{sec:5} in details). The $\Ns=2$ SUSY-SYK model consists of $N$ complex fermions, each of which interacts with $(q-1)$ other fermions with a coupling constant $C_{i_1\dots i_q}$ which is totally antisymmetric with respect to the permutation of indices. These coupling constants are complex independent Gaussian random variables with zero mean and variance
\begin{align}
  \label{eq:2.2.26}
  \overline{|C_{i_1\dots i_q}|^2} = \frac{J^2(q-1)!}{N^{q-1}}.
\end{align}
Similar to the $\Ns=1$ case, $|C_{i_1\dots i_q}|^2$ and $J^2$ have the unit of energy. The Lagrangian is
\begin{align}
  \label{eq:2.22}
  \begin{aligned}
    \mathcal L = \int d\bar\theta \bar\Psi_i D \Psi_i + \frac{i^{\frac{q-1}{2}}}{q!} \left( \int d\theta C_{i_1\dots i_q} \Psi_{i_1}\dots \Psi_{i_q} + \int d\bar\theta \bar C_{i_1\dots i_q} \bar\Psi_{i_1}\dots \bar\Psi_{i_q}\right).
  \end{aligned}
\end{align}
Here the superfield
\begin{align}
  \label{eq:2.23}
  \Psi_i &= \Psi_i(t,\theta) \equiv \psi_i(t) + \theta b_i(t),\\
  \bar\Psi_i &= \bar\Psi_i(t,\bar\theta) \equiv \bar\psi_i(t) + \bar\theta \bar b(t)
\end{align}
represent the complex fermion $i$; it is defined over the supersymmetric spacetime, which has one time direction, $t$, as before but has two Grassmann direction, $\theta,\bar\theta$. $D= \partial_\theta - i\bar\theta \partial_{t}$ is the covariant derivative. The factor $i^{\frac{q-1}{2}}$ ensures the reality of the Lagrangian. The factor $q!$ again arises from the antisymmetry of $C$.

The double commutator of this model is defined as
\begin{align}
  \label{eq:2.25}
  &\frac{1}{N} F(s_1,\theta_1;\bar s_2,\bar\theta_2) \equiv \frac{1}{N^2} \sum_{i,j} \overline{\langle \Phi_i(0,0)\Psi_j(is_1,\theta_1) \bar\Phi_i(\frac{\beta}{2},0) \bar\Psi_j(\frac{\beta}{2} + i\bar s_2,\bar\theta_2) \rangle_{\mathcal C}}.
\end{align}
Here $s_i \equiv t_i-i\theta_i\bar\theta_i, \bar s_i \equiv t_i+i\theta_i\bar\theta_i$ with $t\in\mathbb R^+$ and $\theta_i,\bar\theta_i$ being the Grassmann coordinate, and $\langle\dots\rangle_{\mathcal C}$ denotes the functional average on the contour $\mathcal C$ in Fig. \ref{fig:0}, i.e.
\begin{align}
  \label{eq:2.26}
  \langle\dots\rangle_{\mathcal C} \equiv \frac{\int\mathscr D [\{\Psi_i\},\{\bar\Psi_i\}] e^{\int_\mathcal C dt \mathcal L} (\dots) }{\int\mathscr D [\{\Psi_i\},\{\bar\Psi_i\}] e^{\int_\mathcal C dt \mathcal L}}.
\end{align}
Moreover, $\bar\Phi_i$ is defined in the same way as Eq.(\ref{eq:2.6.1}).

Similar to the derivation of Eq.(\ref{eq:2.8}), we can obtain the following Bethe-Salpeter equation
\begin{align}
  \label{eq:2.30}
  \begin{aligned}
    F(s_1, \theta_1;\bar s_2,\bar\theta_2) &= F_0(s_1, \theta_1,\bar s_2,\bar\theta_2)\\
    &+ \int ds_4 d\theta_4 d\bar s_3 d\bar\theta_3 K_R(s_1, \theta_1;\bar s_2,\bar\theta_2|\bar s_3, \bar \theta_3;s_4, \theta_4) F(\bar s_3, \bar \theta_3;s_4, \theta_4),
  \end{aligned}
\end{align}
for $N\gg 1$, where the retarded kernel
\begin{align}
  \label{eq:2.29}
  \begin{aligned}
    K_R(s_1, \theta_1;\bar s_2, \bar\theta_2&|\bar s_3, \bar\theta_3;s_4, \theta_4)\\
    &= J^2(q-1) G_R(s_1, \theta_1|\bar s_3, \bar\theta_3)G_R(\bar s_2, \bar\theta_2|s_4, \theta_4)(G_W(\bar s_3, \bar\theta_3;s_4, \theta_4))^{q-2}.
  \end{aligned}
\end{align}
Equations (\ref{eq:2.30}) and (\ref{eq:2.29}) hold for general $q$.

For $q\gg 1$, the Wightman function is
\begin{align}
  \label{eq:2.33}
  G_W(\bar s_3, \bar\theta_3;s_4, \theta_4) = \frac{1}{4}\left(1+\frac{1}{q} \ln \frac{\pi v/(\beta \mathcal J)}{\cosh \frac{\pi v (\bar s_3-s_4)}{\beta} + \frac{2\pi v}{\beta} \bar\theta_3\theta_4}\right),
\end{align}
with
\begin{align}
  \label{eq:2.33.1}
  \mathcal J = \frac{qJ^2}{2^{2q-3}}
\end{align}
and $v$ again satisfies Eq.(\ref{eq:0.2}). Furthermore, the retarded Green function $G_R(s_1, \theta_1|\bar s_3, \bar\theta_3)$ can be replaced by the free one, $G_{R0}(s_1, \theta_1|\bar s_3, \bar\theta_3) = -\frac{i}{2}\Theta(s_1-\bar s_3 + 2i\theta_1\bar\theta_3)$. As a result, Eq.(\ref{eq:2.29}) is simplified as
\begin{align}
  \label{eq:2.34}
  K_R(s_1, \theta_1;\bar s_2, \bar\theta_2|\bar s_3, \bar\theta_3;s_4, \theta_4) &= -\frac{1}{4}\Theta(s_1-\bar s_3+i\theta_1\theta_3) \Theta(\bar s_2-s_4+i\theta_2\theta_4) W(\bar s_3, \bar\theta_3;s_4, \theta_4),
\end{align}
where
\begin{align}
  \label{eq:2.35}
  W(\bar s_3, \bar\theta_3;s_4, \theta_4) \equiv J^2(q-1)(G_W(\bar s_3, \bar\theta_3;s_4, \theta_4))^{q-2} = \frac{2\pi v}{\beta} \frac{1}{\cosh\frac{\pi v(\bar s_3-s_4)}{\beta} + \frac{2\pi v}{\beta}\bar\theta_3\theta_4}.
\end{align}

Similar to the $\Ns=1$ case, we can ignore the first term on the right-hand side of Eq.(\ref{eq:2.30}). By performing the derivative: $D_1\bar D_2$ on both sides, where $D_1$ acts on the arguments $s_1,\theta_1$ and $\bar D_2 \equiv \partial_{\theta_2} + i\theta_2 \partial_{t_2}$ on $\bar s_2,\bar\theta_2$, we cast Eq.(\ref{eq:2.30}) into
\begin{align}
  \label{eq:2.36}
  D_1\bar D_2 F(s_1,\theta_1;\bar s_2,\bar\theta_2) = W(\bar s_1,\bar\theta_1;s_2,\theta_2) F(\bar s_1,\bar\theta_1;s_2,\theta_2).
\end{align}
Comparing this equation with Eq.(\ref{eq:2.15}), we find that the arguments on its left- and right-hand sides are different. This difference inherits from the structure of Eq.(\ref{eq:2.30}). To solve this equation we first perform the time translation:
\begin{align}
  \label{eq:2.37}
  \left(\!\!
    \begin{array}{c}
      W(s_1,\theta_1;\bar s_2,\bar\theta_2) \\
      F(s_1,\theta_1;\bar s_2,\bar\theta_2) \\
    \end{array}
  \!\!\right)
   \! = \! e^{i(-\theta_1\bar\theta_1 \partial_{t_1} + \theta_2\bar\theta_2 \partial_{t_2})}\left(\!\!
    \begin{array}{c}
      W(t_1,\theta_1;t_2,\bar\theta_2) \\
      F(t_1,\theta_1;t_2,\bar\theta_2) \\
    \end{array}
  \!\!\right),\,\,\,\,\,
\end{align}
and likewise for $W(\bar s_1,\bar\theta_1;s_2,\theta_2)$ and $F(\bar s_1,\bar\theta_1;s_2,\theta_2)$. Then, we expand the time-translated $W$ and $F$ in Grassmann variables as
\begin{align}
  \label{eq:2.38}
  \left(\!\!
    \begin{array}{c}
      W(t_1,\theta_1;t_2,\bar\theta_2) \\
      F(t_1,\theta_1;t_2,\bar\theta_2) \\
    \end{array}
  \!\!\right)
   \!\equiv\! \left(\!\!
    \begin{array}{c}
      W^\psi(t_1;t_2)  \\
      F^\psi(t_1;t_2)  \\
    \end{array}
  \!\!\right)\!+\!\theta_1\bar\theta_2 \left(\!\!
    \begin{array}{c}
      W^b(t_1;t_2)  \\
      F^b(t_1;t_2)  \\
    \end{array}
  \!\!\right)\,\,\,\,\,
\end{align}
and
\begin{align}
  \label{eq:2.39}
  \left(\!\!
    \begin{array}{c}
      W(t_1,\theta_1;t_2,\bar\theta_2)  \\
      F(t_1,\theta_1;t_2,\bar\theta_2)  \\
    \end{array}
  \!\!\right)
   \!\equiv\! \left(\!\!
    \begin{array}{c}
      W^\psi(t_1;t_2)  \\
      F^\psi(t_1;t_2)  \\
    \end{array}
  \!\!\right)\!+\!\bar\theta_1\theta_2 \left(\!\!
    \begin{array}{c}
      W^b(t_1;t_2)  \\
      F^b(t_1;t_2)  \\
    \end{array}
  \!\!\right).\,\,\,\,\,
\end{align}
With their substitutions Eq.(\ref{eq:2.35}) reduces to
\begin{align}
  \label{eq:2.40}
  \left(\partial_{t_1}\partial_{t_2} - \frac{1}{4} ((W^\psi)^2-W^b)\right)F^\psi = 0
\end{align}
and
\begin{align}
  \label{eq:2.41}
  F^b = -W^\psi F^\psi.
\end{align}
With the help of the explicit expression of $W^\psi$ and $W^b$ obtained from Eq.(\ref{eq:2.35}), Eq.(\ref{eq:2.40}) exactly reduces to Eq.(\ref{eq:6.45}). Therefore, the solution of Eq.(\ref{eq:2.40}) is given by Eq.(\ref{eq:2.20.1}). Importantly, the chaos exponent is again given by Eqs.(\ref{eq:0.1}) and (\ref{eq:0.2}). Combining the solution for $F^\psi$ with Eq.(\ref{eq:2.41}) and (\ref{eq:2.38}) we obtain $F^b$ and the full $F(t_1,\theta_1;t_2,\bar\theta_2)$. Upon further undoing the time translation we obtain
\begin{align}
  \label{eq:2.42}
  F^b(s_1;\bar s_2) = -\frac{\pi v}{\beta} \frac{e^{\lambda_L \frac{s_1+\bar s_2}{2}}}{\cosh^2\frac{\pi v(s_1-\bar s_2)}{\beta}}
\end{align}
and
\begin{align}
  \label{eq:2.43}
  F(s_1,\theta_2;\bar s_2,\bar\theta_2) = \frac{e^{\lambda_L \frac{s_1+\bar s_2}{2}}}{\cosh \frac{\pi v (s_1-\bar s_2)}{\beta} + \frac{2\pi v}{\beta} \theta_1\bar\theta_2}.
\end{align}

\section{The \texorpdfstring{$\boldsymbol{(N|M)}$}{(N|M)}-SYK model  and its Green functions}
\label{sec:3}

The $(N|M)$-SYK model was introduced in \cite{Marcus18}. It consists of $N$ Majorana fermions $\psi_i$ and $M$ bosons $b_\alpha$. Each boson interacts with $(q-1)$ Majorana fermions, and each fermion interacts with a boson and $(q-2)$ fermions. The coupling constant is $C_{\alpha [i_1\dots i_{q-1}]}$, where $\alpha$ labels the bosons and $i_1, \dots, i_{q-1}$ label the fermions. Importantly, it is antisymmetric only with respect to the fermionic indices $i_1, \dots, i_{q-1}$ in the bracket of the subscript of $C$. The coupling constants $C_{\alpha [i_1 \dots i_{q-1}]}$s are real. They are independent Gaussian random variables with zero mean and variance
\begin{align}
  \label{eq:1.0}
  \overline{C_{\alpha [i_1\dots i_{q-1}]}^2} = \frac{(q-1)! J^2 p}{N^{q-1}}, \quad p = \sqrt{\frac{N}{M}}.
\end{align}
The Lagrangian is
\begin{align}
  \label{eq:1.1}
  \mathcal L = -\psi_i \partial_\tau \psi_i + b_\alpha b_\alpha + \frac{i^\frac{q-1}{2}}{(q-1)!} C_{\alpha [i_1\dots i_{q-1}]} b_\alpha \psi_{i_1} \dots \psi_{i_{q-1}}.
\end{align}
The factor $(q-1)!$ arises from this antisymmetry, and $i^{\frac{q-1}{2}}$ ensures the reality of the Lagrangian. According to Eq.(\ref{eq:1.1}), $C^2_{\alpha i_1\dots i_{q-1}}$ and thereby $J^2$ have the dimension of energy. For $p=1$ Eqs.(\ref{eq:1.0}) and (\ref{eq:1.1}) become the $\Ns=1$ SUSY-SYK model, if $C$ in Eq.(\ref{eq:1.1}) were antisymmetric in all its indices. In \cite{Marcus18}, the $(N|M)$-SYK model in the strong coupling limit was studied. In this work, we address arbitrary interaction strength $J$, but for large $q$.

For the convenience below here we compute the Euclidean two-point fermionic and bosonic correlator, defined as
\begin{align}
  \label{eq:3.1}
  \begin{aligned}
    G^\psi(\tau_1|\tau_2) &\equiv \frac{1}{N} \overline{\langle\psi_i(\tau_1)\psi_i(\tau_2)\rangle_{\mathcal C}},\\
    G^b(\tau_1|\tau_2) &\equiv \frac{p^2}{N} \overline{\langle b_\alpha(\tau_1)b_\alpha(\tau_2)\rangle_{\mathcal C}},
  \end{aligned}
\end{align}
respectively. From now on we consider $N\gg 1$. It can be readily shown that for $q\gg 1$,
\begin{align}
  \label{eq:14}
  &\begin{aligned}
    G^{\psi}(\tau_1|\tau_2) &= G^{\psi}_0(\tau_1|\tau_2) + \int d\tau_a d\tau_b G^\psi_0(\tau_1|\tau_a)\Sigma^\psi(\tau_a|\tau_b)G^\psi(\tau_b|\tau_2),\\
    G^{b}(\tau_1|\tau_2) &= G^{b}_0(\tau_1|\tau_2) + \int d\tau_a d\tau_b G^{b}_0(\tau_1|\tau_a)\Sigma^b(\tau_a|\tau_b)G^{b}(\tau_b|\tau_2),
  \end{aligned}
\end{align}
with the self-energies
\begin{align}
  \label{eq:13}
  &\begin{aligned}
    \Sigma^\psi(\tau_1|\tau_2) &= J^2\frac{q-1}{p} G^b(\tau_1|\tau_2) (G^\psi(\tau_1|\tau_2))^{q-2},\\
    \Sigma^b(\tau_1|\tau_2) &= J^2p(G^\psi(\tau_1|\tau_2))^{q-1},
  \end{aligned}
\end{align}
and the free propagators
\begin{align}
  \label{eq:3.8}
  \begin{aligned}
    G^{\psi}_0(\tau_1|\tau_2) &= \frac{1}{2} \sgn(\tau_1-\tau_2),\\
    G^{b}_0(\tau_1|\tau_2) &= -\delta(\tau_1-\tau_2).
  \end{aligned}
\end{align}
Because of $|G^{\psi,b}|<1$, which will be shown to be satisfied by the explicit expression of $G^{\psi,b}$ derived below, Eq.(\ref{eq:13}) implies that the self-energy vanishes for $q\gg 1$. So, we can expand $G^{\psi,b}$ in $1/q$:
\begin{align}
  \label{eq:15}
  \begin{aligned}
    G^\psi(\tau_1|\tau_2) &= \frac{1}{2}\text{sgn}(\tau_1-\tau_2) \left(1 + \frac{1}{q} g^\psi(\tau_1|\tau_2) + O(\frac{1}{q^2})\right),\\
    G^b(\tau_1|\tau_2) &= -\delta(\tau_1-\tau_2) + \frac{1}{q} g^b(\tau_1|\tau_2) + O(\frac{1}{q^2}).
  \end{aligned}
\end{align}
Combining Eqs.(\ref{eq:14}),(\ref{eq:13}) and (\ref{eq:15}), we obtain
\begin{align}
  \label{eq:16}
  &\begin{aligned}
    \partial_{\tau_1}\partial_{\tau_2} &\big(\text{sgn}(\tau_1-\tau_2) g^\psi(\tau_1|\tau_2)\big) = -\frac{2\mathcal J}{p}\text{sgn}(\tau_1-\tau_2) e^{g^\psi(\tau_1|\tau_2)} g^b(\tau_1|\tau_2),
  \end{aligned}\\
  \label{eq:17}
  &g^b(\tau_1|\tau_2) = -\frac{\mathcal J p}{2} e^{g^\psi(\tau_1|\tau_2)},
\end{align}
where
\begin{align}
  \label{eq:17.1}
  \mathcal J = \frac{qJ^2}{2^{q-2}}
\end{align}
is fixed throughout this and the next section. From Eqs.(\ref{eq:16}) and (\ref{eq:17}) we can see that $g^{\psi,b}(\tau_1|\tau_2) = g^{\psi,b}(\tau_1-\tau_2)$. So the time arguments of $g^{\psi,b}$ will be suppressed below. By substituting Eq.(\ref{eq:17}) into Eq.(\ref{eq:16}), we obtain
\begin{align}
  \label{eq:18}
  \partial_{\tau}^2 \big(\text{sgn}(\tau) g^\psi\big) -\mathcal J^2 \text{sgn}(\tau) e^{2g^\psi} = 0.
\end{align}
In Appendix \ref{app:1}, we solve this equation and find its solution to be
\begin{align}
  \label{eq:19}
  e^{g^\psi} &= \frac{\pi v}{\beta \mathcal J} \frac{1}{\cos \left( \pi v \left(\frac{|\tau_1-\tau_2|}{\beta} - \frac{1}{2}\right)\right)},
\end{align}
where $v$ satisfies Eq.(\ref{eq:0.2}) with $x=\beta\mathcal J$. With its substitution into Eq.(\ref{eq:17}) we obtain
\begin{align}
  \label{eq:20}
  g^b &= -\frac{\pi v p}{2\beta} \frac{1}{\cos\left( \pi v \left(\frac{|\tau_1-\tau_2|}{\beta} - \frac{1}{2}\right)\right)}.
\end{align}
It is important that for any finite $x>0$ Eq.(\ref{eq:0.2}) always has a solution in the interval $(0,1)$. However, depending on $x$ more solutions can result. When this occurs, we choose the solution $v$ satisfying $v \in (0,1)$, so that Eq.(\ref{eq:19}) remains finite for any $|\tau_1-\tau_2|\in [0, \beta]$. Then, from the right-hand side of Eq.(\ref{eq:19}) we find that $\ln\frac{\pi v}{\beta \mathcal J} \leq g^\psi \leq 0$. Therefore, for $q\gg 1$ and finite $\beta\mathcal J$ we have $\frac{|g^\psi|}{q} \ll 1$, which implies $|G^\psi|<1$ according to the first equation of Eq.(\ref{eq:15}). Likewise, taking Eq.(\ref{eq:20}) into account we have $|G^b|<1$ (for $\tau_1\neq \tau_2$). So the validity of the large $q$-expansion of $G^{\psi,b}$ is justified.

Furthermore, we calculate the Wightman functions, which are defined as
\begin{align}
  \label{eq:2.C.14}
  G^\psi_W(t_1;t_2) &\equiv G^\psi(it_1+\frac{\beta}{2}|it_2),\\
  \label{eq:2.C.15}
  G^b_W(t_1;t_2) &\equiv G^b(it_1+\frac{\beta}{2}|it_2).
\end{align}
With the substitution of Eqs.(\ref{eq:15}) and (\ref{eq:19}), we obtain
\begin{align}
  \label{eq:2.C.17}
  G^\psi_W(t_1;t_2) &= \frac{1}{2}\left(1 + \frac{1}{q} \ln \frac{\pi v/(\beta \mathcal J)}{\cosh\frac{\pi v (t_1-t_2)}{\beta}}\right),\\
  \label{eq:2.C.18}
  G^b_W(t_1;t_2) &= -\frac{\pi v p}{2\beta q} \frac{1}{\cosh \frac{\pi v (t_1-t_2)}{\beta}}.
\end{align}

\section{Non-maximal chaos in the \texorpdfstring{$\boldsymbol{(N|M)}$}{(N|M)}-SYK model}
\label{sec:4}

For the $(N|M)$-SYK model, the fermionic double commutator is defined as
\begin{align}
  \label{eq:1.2}
  \begin{aligned}
    &\frac{1}{N} F^\psi(t_1;t_2) \equiv \frac{1}{N^2} \overline{\langle \phi_i(0) \psi_j(it_1) \phi_i(\frac{\beta}{2}) \psi_j(\frac{\beta}{2}+it_2)\rangle_{\mathcal C}},
  \end{aligned}
\end{align}
where
\begin{align}
  \label{eq:1.2.1}
  \phi_i(\tau) \equiv \psi_i(\tau^+) - \psi_i(\tau^-)
\end{align}
and $\langle\dots\rangle_{\mathcal C}$ denotes the functional average on the contour $\mathcal C$ shown in Fig.\ref{fig:0}, i.e.
\begin{align}
  \label{eq:1.21}
  \langle\dots\rangle_{\mathcal C} \equiv \frac{\int\mathscr D [\{\psi_i\}]\mathscr D [\{b_\alpha\}] e^{\int_\mathcal C dt \mathcal L} (\dots) }{\int\mathscr D [\{\psi_i\}]\mathscr D [\{b_\alpha\}] e^{\int_\mathcal C dt \mathcal L}}.
\end{align}
The bosonic double commutator is defined as
\begin{align}
  \label{eq:1.3}
  \begin{aligned}
    &\frac{1}{N} F^b(t_1;t_2) \equiv \frac{p^2}{N^2} \overline{\langle \phi_i(0) b_\alpha(it_1) \phi_i(\frac{\beta}{2}) b_\alpha(\frac{\beta}{2}+it_2)\rangle_{\mathcal C}}.
  \end{aligned}
\end{align}In this section, we calculate Eqs.(\ref{eq:1.2}) and (\ref{eq:1.3}) by generalizing the calculations in Sec. \ref{sec:2.1}.

\subsection{The Bethe-Salpeter equation for the double commutator}
\label{sec:4.1}

\begin{figure}[b]
  \centering
  \includegraphics[width=\textwidth]{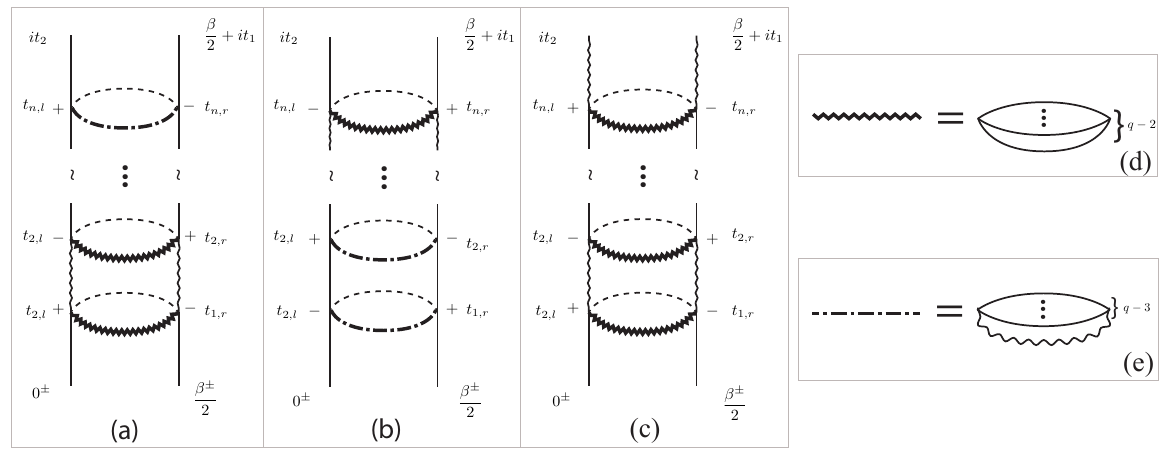}
  \caption{Examples of the diagrams for the fermionic double commutator (a,b) and the bosonic double commutator (c). Each diagram is composed of two vertical rails and many rungs. Each rung consists of $(q-2)$ fermionic Wightman propagators (d) or $(q-3)$ fermionic Wightman propagators, a bosonic Wightman propagator (e) and an interaction line.}
  \label{fig:1}
\end{figure}

Similar to the $\Ns=1$ SUSY-SYK model (cf. Fig. \ref{fig:6.1}), for $N\gg 1$ the diagrams dominating the fermionic and bosonic commutator are exemplified by Fig. \ref{fig:1}(a)-(c), each of which has two vertical rails and many horizontal rungs. Like in Fig. \ref{fig:6.1}, the complex times of the vertexes on the left (right) rail are located on the left (right) rail of the contour $\mathcal C$ in Fig. \ref{fig:0}. Moreover, the imaginary part of the complex time of a vertex increases, when the vertex moves from the bottom of a rail to the top. However, unlike in Fig. \ref{fig:6.1}, each rail is now made up of fermionic propagators $G^\psi$ (thin solid line) and bosonic propagators $G^b$ (thin wavy line). For the fermionic (bosonic) double commutator, a fermionic (bosonic) propagator leads the rails. Correspondingly, there are two types of rungs: One type of rungs is the product (heavy zigzag line) of $(q-2)$ fermionic Wightman functions $G^\psi_W$ analytically continued from $G^\psi$ [Fig. \ref{fig:1}(d)], and an interaction line (dashed line); the other is the product (dash-dotted line) of $(q-3)$ fermionic $G^\psi_W$, one bosonic Wightman function $G^b_W$ analytically continued from $G^b$ [Fig. \ref{fig:1}(e)], and an interaction line.

Owing to the above important differences of diagrammatic structures, we find that upon summing up $n$-rung diagrams for $F^\psi$ ($F^b$) of different $+/-$ configurations in each rail, a diagram, denoted as $F^\psi_n$ ($F^b_n$), results, but the diagrams representing $F^{\psi,b}_{n+1}$ are generated from those representing $F^{\psi,b}_n$ in a way quite different from that described by Eq.(\ref{eq:6.3}). To be specific, $F^\psi_{n+1}$ can be generated from either $F^\psi_n$ or $F^b_n$ according to
\begin{align}
  \label{eq:4.3}
  F^\psi_{n+1}(t_1;t_2) = \int &dt_a dt_b \Big(K_{11}(t_1;t_2|t_a;t_b)F^\psi_{n}(t_a;t_b) + K_{12}(t_1;t_2|t_a;t_b)F^b_{n}(t_a;t_b) \Big)
\end{align}
with the corresponding retarded kernel being
\begin{align}
  \label{eq:2}
  K_{11}&(t_1;t_2|t_a;t_b) = J^2\frac{(q-1)(q-2)}{p} G_{R}^\psi(t_1|t_a) G_{R}^\psi(t_2|t_b) G_{W}^b(t_a;t_b) (G_{W}^\psi(t_a;t_b))^{q-3}
\end{align}
and
\begin{align}
  \label{eq:3}
  K_{12}&(t_1;t_2|t_a;t_b) = -J^2\frac{(q-1)}{p} G_{R}^\psi(t_1|t_a) G_{R}^\psi(t_2|t_b) (G_{W}^\psi(t_a;t_b))^{q-2}.
\end{align}
For $F^b_{n+1}$, it can be generated only from $F^\psi_n$ according to
\begin{align}
  \label{eq:4.4}
  F^b_{n+1}(t_1;t_2) = \int dt_a dt_b K_{21}(t_1;t_2|t_a;t_b)F^\psi_{n}(t_a;t_b)
\end{align}
with the retarded kernel
\begin{align}
  \label{eq:4}
  K_{21}&(t_1;t_2|t_a;t_b) = J^2(q-1)p G_{R}^b(t_1|t_a) G_{R}^b(t_2|t_b) (G_{W}^\psi(t_a;t_b))^{q-2}.
\end{align}
In Eqs.(\ref{eq:2})-(\ref{eq:4}),
\begin{align}
  \label{eq:4.1}
  G^{\psi,b}_R(t_1|t_2) = -i\Theta(t_1-t_2) \Big(G^{\psi,b}(it_1|it_2-\epsilon) - G^{\psi,b}(it_1|it_2+\epsilon)\Big),
\end{align}
are the fermionic (bosonic) retarded Green function, and
\begin{align}
  \label{eq:4.2}
    G^{\psi,b}_{W}(t_1;t_2) = G^{\psi,b}(it_1+\frac{\beta}{2}|it_2).
\end{align}
are the fermionic (bosonic) Wightman function. Note that they are all obtained from $G^{\psi,b}(t_1|t_2)$ ($t_{1,2}\in\mathbb R$) through analytic continuation.

Comparing Eqs.(\ref{eq:4.3})-(\ref{eq:4}) with Eqs.(\ref{eq:2.12.1})-(\ref{eq:2.16.1}), we find that $K_{11}|_{p=1} = K_R^{(11)}$, $K_{12}|_{p=1} = -K_R^{(12)}$ and $K_{21}|_{p=1} = -K_R^{(21)}$. Note that the minus signs for the $12$- and $21$-sector arise from different conventions in defining $F^b$. More precisely, if in Eq.(\ref{eq:2.16}) $F^b$ is expanded in $\theta_2\theta_1$ instead of $\theta_1\theta_2$, then the minus sign disappear. The ensuing argument is not surprising, because we have shown above that for $p=1$ the $(N|M)$-SYK model recovers the $\Ns=1$ SUSY-SYK model.

Noting that $F^{\psi,b} = \sum_{n=0}^\infty F^{\psi,b}_{n}$, with the help of Eqs.(\ref{eq:4.3}) and (\ref{eq:4.4}) we obtain the Bethe-Salpeter equation:
\begin{align}
  \label{eq:4.5}
  &\begin{aligned}
    F^\psi(t_1;t_2) = &F^\psi_{0}(t_1;t_2) + \int dt_a dt_b \Big(K_{11}(t_1;t_2|t_a;t_b)F^\psi(t_a;t_b) + K_{12}(t_1;t_2|t_a;t_b)F^b(t_a;t_b) \Big)
  \end{aligned}
\end{align}
for the fermionic double commutator and
\begin{align}
  \label{eq:4.6}
  &\begin{aligned}
    F^b(t_1;t_2) = &F^b_{0} (t_1;t_2) + \int dt_a dt_b K_{21}(t_1;t_2|t_a;t_b)F^\psi(t_a;t_b)
  \end{aligned}
\end{align}
for the bosonic double commutator.

\subsection{Solution of the Bethe-Salpeter equation}
\label{sec:4.2}

For $q\gg 1$, we can simplify Eqs.(\ref{eq:2})-(\ref{eq:4}) as
\begin{align}
  \label{eq:5.24}
  K_{11}(t_1;t_2|t_a;t_b) &=\frac{1}{p} G_{R0}^\psi(t_1|t_a)G_{R0}^\psi(t_2|t_b)W^b(t_a;t_b),\\
  \label{eq:5.25}
  K_{12}(t_1;t_2|t_a;t_b) &=-\frac{1}{p} G_{R0}^\psi(t_1|t_a)G_{R0}^\psi(t_2|t_b)W^\psi(t_a;t_b),\\
  \label{eq:5.26}
  K_{21}(t_1;t_2|t_a;t_b) &=p G_{R0}^b(t_1|t_a)G_{R0}^b(t_2|t_b)W^\psi(t_a;t_b),
\end{align}
where
\begin{align}
  \label{eq:5.27}
  \begin{aligned}
    G_{R0}^\psi(t_1|t_2) &= -i\Theta(t_1-t_2),\\
    G_{R0}^b(t_1|t_2) &= -\delta(t_1-t_2),
  \end{aligned}
\end{align}
are free retarded propagators and
\begin{align}
  \label{eq:2.C.20}
  W^\psi(t_1;t_2) &\equiv J^2 (q-1) (G_W^\psi(t_1;t_2))^{q-2} = \frac{\pi v}{\beta} \frac{1}{\cosh \frac{\pi v (t_1-t_2)}{\beta}}
\end{align}
and
\begin{align}
  \label{eq:2.C.21}
  W^b(t_1;t_2) &\equiv J^2 (q-1) (q-2) G_W^b(t_1;t_2) (G_W^\psi(t_1;t_2))^{q-3} = - \frac{p\pi^2 v^2 }{\beta^2} \frac{1}{\cosh^2 \frac{\pi v (t_1-t_2)}{\beta}}.
\end{align}
By substituting Eqs.(\ref{eq:5.24}) and (\ref{eq:5.25}) into Eq.(\ref{eq:4.5}) and taking the derivative $-\partial_{t_1}\partial_{t_2}$ on both sides, we obtain
\begin{align}
  \label{eq:5.31}
  \partial_{t_1}\partial_{t_2} F^\psi &=  \frac{1}{p} \left(W^\psi F^b - W^b F^\psi\right).
\end{align}
By substituting Eq.(\ref{eq:5.26}) into Eq.(\ref{eq:4.6}), we obtain
\begin{align}
  \label{eq:5.32}
  F^b &= pW^\psi F^\psi.
\end{align}
Then we substitute Eq.(\ref{eq:5.32}) into Eq.(\ref{eq:5.31}), obtaining
\begin{align}
  \label{eq:8}
  \left[\partial_{t_1}\partial_{t_2} - \big((W^\psi)^2 -\frac{1}{p} W^b\big)\right] F^\psi= 0.
\end{align}
With the substitution of Eqs.(\ref{eq:2.C.20}) and (\ref{eq:2.C.21}) we can write it exactly as Eq.(\ref{eq:6.45}), whose solution is given by Eq.(\ref{eq:2.20.1}). Once more, the chaos exponent follows the single-parameter scaling law Eqs.(\ref{eq:0.1}) and (\ref{eq:0.2}). Substituting the solution into Eq.(\ref{eq:5.32}) we obtain
\begin{align}
  \label{eq:25}
  F^b(t_1;t_2) = \frac{p\pi v}{\beta}\frac{e^{\lambda_L \frac{t_1+t_2}{2}}}{\cosh^2 \frac{\pi v (t_1-t_2)}{\beta}}.
\end{align}
In sharp contrast to Eq.(\ref{eq:25}), which depends on the value of $p$, the universality exhibited by $\lambda_L$ is completely independent of $p$, and in particular, is regardless of whether the model is supersymmetric ($p=1$) or not ($p\neq 1$). 

\section{Non-maximal chaos in more general SYK-like models}
\label{sec:5}

We have seen that the chaos exponent of the $\Ns=0,1,2$ SUSY- and the $(N|M)$-SYK model all satisfy the scaling law described by Eqs.(\ref{eq:0.1}) and (\ref{eq:0.2}), when the interaction strength is arbitrary. It is natural to ask whether this scaling law may apply to other 1D variants of the SYK model. In this section, by refining studies in Secs. \ref{sec:2} - \ref{sec:4}, we make a conjecture that provides an answer to this question.

First of all, we observe that although the double commutator of the $\Ns=1,2$ SUSY- and the $(N|M)$-SYK model all have two components, $F^\psi(t_1;t_2)$ and $F^b(t_1;t_2)$, chaotic behaviors are completely encoded in $F^\psi(t_1;t_2)$. In particular, Eqs.(\ref{eq:2.17}), (\ref{eq:2.40}) and (\ref{eq:8}) all reduce to
\begin{align}
  \label{eq:5.1}
  \left[\partial_{t_1}\partial_{t_2} - \frac{2\pi^2}{\beta^2} \frac{1}{\cosh^2 \frac{\pi (t_1-t_2)}{\beta}}\right] F^\psi(t_1;t_2) = 0
\end{align}
in the strong coupling limit $\beta\mathcal J \rightarrow \infty$, which is the special case of Eq.(\ref{eq:6.45}) for $v=1$. and has been derived for the SYK model \cite{Maldacena16}. The solution of Eq.(\ref{eq:5.1}),
\begin{align}
  \label{eq:5.2}
  F^\psi(t_1;t_2) \sim e^{\lambda_L (t_1+t_2)/2}, \quad \lambda_L = \frac{2\pi}{\beta}
\end{align}
then describes the maximal chaoticity.

What happens away from the strong coupling limit? That Eqs.(\ref{eq:6.45}), (\ref{eq:2.40}) and (\ref{eq:8}) all can reduce to Eq.(\ref{eq:6.45}) suggests a simple scenario. That is, we just need to rescale the time in Eq.(\ref{eq:5.1}) by an appropriate parameter $v$:
\begin{align}
  \label{eq:5.3}
  t_{1,2} \rightarrow vt_{1,2},
\end{align} 
and insert it into the solution of the ensuing equation, namely, Eq.(\ref{eq:5.2}). Consequently, the chaos exponent differs from the maximal chaos bound by a factor of $v$, i.e.,
\begin{align}
  \label{eq:5.4}
  \frac{2\pi}{\beta} \rightarrow \frac{2\pi v}{\beta} = \lambda_L.
\end{align}

The reminder is to determine $v$ in Eq.(\ref{eq:5.4}). To this end we recall that, irrespective of the models considered, in the strong coupling limit the Euclidean propagator of fermions takes a general form:
\begin{align}
  \label{eq:5.5}
  G^\psi(\tau) = A\,\sgn(\tau)\left(\frac{\pi}{\beta \mathcal J \cos\left(\pi(\frac{1}{2} - \frac{|\tau|}{\beta})\right)}\right)^{\Delta_\psi},
\end{align}
owing to the emergence of conformal symmetry. Here $A$ is a numerical factor involved in the definition of $G^\psi$ and is model-dependent. For example, $A=\frac{1}{2}$ for the SYK model , the $\Ns=1$ SUSY- and the $(N|M)$-SYK model, while $A=\frac{1}{4}$ for the $\Ns=2$ SUSY-SYK model. The power $\Delta_\psi$ and the expression of $\mathcal J$ in terms of $J$ and $q$ are model-dependent also. Corresponding to the rescaling (\ref{eq:5.3}), for arbitrary $\beta\mathcal J$ Eq.(\ref{eq:5.5}) is modified as
\begin{align}
  \label{eq:5.6}
  G^\psi(\tau) \rightarrow A\,\sgn(\tau)\left(\frac{\pi v}{\beta \mathcal J \cos\left(\pi v(\frac{1}{2} - \frac{|\tau|}{\beta})\right)}\right)^{\Delta_\psi}.
\end{align}
Because this modification cannot alter the value of $G^\psi$ at $\tau=0$, we have
\begin{align}
  \label{eq:5.7}
  \frac{\pi v}{\beta \mathcal J \cos \frac{\pi v}{2}} = 1.
\end{align}
This gives Eq.(\ref{eq:0.2}).

So, we find that, although the scaling law described by Eqs.(\ref{eq:0.1}) and (\ref{eq:0.2}) is for arbitrary interaction strength, it can be attributed to the emergence of conformal symmetry in the strong coupling limit. In other words, the non-maximal chaoticity or integrability for finite interaction strength has intimate relations to the maximal chaoticity in the strong coupling limit. We are not aware of any similar phenomena in chaotic dynamics.

The quantitative arguments above rely little on detailed modifications of the SYK model: Only to be 1D and have large $q$ is the modified model required. We are thus led to the following conjecture: Such modified SYK models  would exhibit non-maximal chaos described by the scaling law Eqs.(\ref{eq:0.1}) and (\ref{eq:0.2}). While the scaling law implies that, as $\beta \mathcal J$ decreases from the infinite, $\lambda_L$ decreases monotonically from the maximal chaos bound to zero, it points out a universal route: At one end of the route quantum motion is maximally chaotic, and at the other is completely regular, namely, integrable.

\section{Concluding remarks}
\label{sec:7}
In this work we have studied non-maximal chaos in SUSY- and $(N|M)$-SYK models. We have calculated the double commutator for the $\Ns=1,2$ SUSY- and the $(N|M)$-SYK model, and shown that their chaos exponents $\lambda_L$ all follow the single-parameter scaling law described by Eqs. (\ref{eq:0.1}) and (\ref{eq:0.2}) for large $q$. We have also found that, notwithstanding weak or strong breaking of conformal symmetry for finite interaction strength, the scaling law is deeply rooted at the emergent conformal symmetry in the strong coupling limit. This leads us to conjecture that the scaling law might hold for other 1D variants of the SYK model, and points out a route from maximal chaoticity to integrability. It would be important to prove/disprove this conjecture in the future or, at least, check explicitly its validity for other models, notably, the $\Ns=4$ SUSY-SYK model \cite{Gates21,Gates22}.

We have seen that the scaling law and the conjecture above hold only in 1D. Indeed, the quantitative arguments made in Sec. \ref{sec:5} cease to work in two dimension (2D). It has been found that 2D variants of the SYK model, e.g., the 2D generalization of $(N|M)$-SYK model \cite{Kim21}, can also exhibit maximal chaos in the strong coupling limit. For such variants we expect similar scaling behaviors to follow away from that limit.

Note that for strong coupling $\beta\mathcal J\gg 1$, Eqs.(\ref{eq:0.1}) and (\ref{eq:0.2}) yield a $\frac{1}{\beta\mathcal J}$-expansion for $\lambda_L$. The first term, which gives the maximal chaos bound, can be interpreted in term of gravitational scattering between two classical point particles \cite{Kitaev15,Maldacena16b}. For the SYK model the second term has been interpreted as the stringy correction to the graviton spin \cite{Maldacena16,Stanford15}. In the future it is interesting to study the gravity interpretation of higher order corrections, and in particular, their universality.

\acknowledgments
This work is supported by NSFC project Nos. 11925507 and 12047503.

\appendix

\section{The solution of Eq.(\ref{eq:6.45})}
\label{app:2}

The solution of Eq.(\ref{eq:6.45}) was given in \cite{Maldacena16} without derivations. In this appendix we provide the derivations.

By the rescaling:
\begin{align}
  \label{eq:B1}
  t_{1,2} \rightarrow \frac{\pi v}{\beta} t_{1,2} \equiv \tilde t_{1,2},
\end{align}
we rewrite Eq.(\ref{eq:6.45}) as
\begin{align}
  \label{eq:B2}
  \left[\partial_{\tilde t_1} \partial_{\tilde t_2} - \frac{2}{\cosh^2 (\tilde t_1 - \tilde t_2)}\right] F^\psi = 0.
\end{align}
Then, we make the following transformation:
\begin{align}
  \label{eq:B3}
  \tilde t \equiv \frac{\tilde t_1 + \tilde t_2}{2},\quad \tilde \tau \equiv \tilde t_1-\tilde t_2
\end{align}
to rewrite Eq.(\ref{eq:B2}) as
\begin{align}
  \label{eq:B4}
  \left(\frac{1}{4} \partial_{\tilde t}^2 - \partial_{\tilde \tau}^2 - \frac{2}{\cosh^2 \tilde \tau}\right) F^\psi = 0.
\end{align}
This equation can be solved by separating variables, i.e.
\begin{align}
  \label{eq:B5}
  F^\psi = e^{\tilde \lambda_L \tilde t} \tilde F^\psi(\tilde \tau).
\end{align}
Inserting it into Eq.(\ref{eq:B4}) gives
\begin{align}
  \label{eq:B6}
  \left[\left(\frac{\tilde \lambda_L}{2}\right)^2 - \frac{d^2}{d\tilde \tau^2} - \frac{2}{\cosh^2 \tilde \tau}\right] \tilde F^\psi = 0.
\end{align}
Making the following transformation:
\begin{align}
  \label{eq:B7}
  x = \sqrt{1-\cosh^{-2}\tilde\tau},
\end{align}
where $0\leq x \leq 1$, we can rewrite Eq.(\ref{eq:B6}) as
\begin{align}
  \label{eq:B8}
  \left[(1-x^2)\partial_x^2 - 2x\partial_x + \left(2 - \frac{(\tilde\lambda_L/2)^2}{1-x^2}\right)\right]\tilde F^\psi = 0.
\end{align}
The solution is the associated Legendre function,
\begin{align}
  \label{eq:B9}
  \tilde F^\psi \propto P_1^{\tilde \lambda_L/2}(x).
\end{align}
For the convenience of discussions below, we rewrite this solution in an equivalent form, read
\begin{align}
  \label{eq:B10}
  \tilde F^\psi(x) \propto \frac{1}{\Gamma\left(1-\frac{\tilde \lambda_L}{2}\right)} \left(\frac{1+x}{1-x}\right)^{\frac{\tilde\lambda_L/2}{2}} {}_2F_1(-1,2;1-\tilde\lambda_L/2;\frac{1-x}{2}),
\end{align}
where ${}_2F_1$ is the hypergeometric function.

Because $\tilde F^\psi$ is regular for $\tilde \tau\rightarrow\infty$, upon making the transformation Eq.(\ref{eq:B7}) $\tilde F^\psi$ is regular at $x=1$. So, by Eq.(\ref{eq:B10}) we have $\tilde \lambda_L/2 \in \mathbb N \cup \{0\}$; otherwise $\tilde F^\psi(x=1)$ diverges. If $\tilde\lambda_L/2 = 0$, then from Eq.(\ref{eq:B5}) we see that no chaotic behaviors arise. Therefore, this solution is not relevant to chaos. In fact, it gives an extended eigenstate of Eq.(\ref{eq:B4}). So, we need to consider only the case of $\tilde\lambda_L/2 \in\mathbb N$. If $\tilde\lambda_L/2 \geq 2$, then we have \cite{Gradshteyn}
\begin{align}
  \label{eq:B11}
  \frac{1}{\Gamma\left(1-\frac{\tilde \lambda_L}{2}\right)} {}_2 F_1(-1,2;1-\frac{\tilde\lambda_L}{2};\frac{1-x}{2}) = 0,
\end{align}
and the solution for $\tilde F^\psi$ is trivial. Therefore, the unique nontrivial solution corresponds to
\begin{align}
  \label{eq:B12}
  \tilde\lambda_L/2 = 1.
\end{align}
Substituting it into Eq.(\ref{eq:B9}) gives
\begin{align}
  \label{eq:B13}
  \tilde F^\psi(x) \propto P_1^1(x) = -\sqrt{1-x^2},
\end{align}
which is the unique bound state of Eq.(\ref{eq:B4}). Combining Eqs.(\ref{eq:B5}), (\ref{eq:B7}), (\ref{eq:B12}) and (\ref{eq:B13}), we obtain
\begin{align}
  \label{eq:B14}
  F^\psi(\tilde t, \tilde \tau) = \frac{e^{2\tilde t}}{\cosh \tilde \tau},
\end{align}
where without loss of generality the proportionality coefficient in Eq.(\ref{eq:B13}) has been determined in the way so that $\tilde F^\psi(0) = 1$. Finally, we substitute Eq.(\ref{eq:B3}) into Eq.(\ref{eq:B14}) and undo the rescaling Eq.(\ref{eq:B1}). As a result, we obtain Eq.(\ref{eq:2.20.1}).

\section{The solution of Eq.(\ref{eq:18})}
\label{app:1}

An equation similar to Eq.(\ref{eq:18}) was reported in \cite{Maldacena16}, and its solution was given there without derivations. For the self-contained purposes in this appendix we provide the details of solving this equation.

By definition $G^\psi = G^\psi_0$ at $\tau= 0$. So we impose the boundary condition: $g^\psi(0) = g^\psi(\pm\beta) = 0$. Define $f \equiv e^{g^\psi}$ and $h \equiv \frac{1}{f}\frac{df}{d\tau}$. With their substitution into Eq.(\ref{eq:18}) we obtain
\begin{align}
  \label{eq:A1}
  h dh= \mathcal J^2 f df.
\end{align}
Integrating Eq.(\ref{eq:A1}) we obtain
\begin{align}
  \label{eq:A2}
  \frac{1}{f} \frac{df}{d\tau} = \sqrt{-\frac{\pi^2 v^2}{\beta^2} + \mathcal J^2 f^2},
\end{align}
where $v$ is real because $f$ is periodic, inheriting from the boundary condition of $g^\psi$, and is to be determined from the boundary condition. Integrating Eq.(\ref{eq:A2}) we obtain
\begin{align}
  \label{eq:A3}
  f = \frac{\pi v}{\beta \mathcal J} \frac{1}{\cos \left(\pi v \left(\frac{\tau}{\beta} - \phi_0\right)\right)},
\end{align}
where $\phi_0$ is to be determined from the boundary condition.

At $\tau = 0$ and $\beta$, Eq.(\ref{eq:A3}) reduces to
\begin{align}
  \label{eq:A4}
  &\frac{\pi v}{\beta \mathcal J} \frac{1}{\cos (\pi v \phi_0)} = 1,
\end{align}
and
\begin{align}
  \label{eq:A5}
  &\frac{\pi v}{\beta \mathcal J} \frac{1}{\cos (\pi v (1-\phi_0))} = 1,
\end{align}
respectively. They imply $\pi v \phi_0 =\pi v (1 - \phi_0)$ because of $f < \infty$ in the interval $[0, \beta]$, which gives $\phi_0 = 1/2$. Substituting this into Eq.(\ref{eq:A4}), we obtain Eq.(\ref{eq:0.2}) with $x=\beta\mathcal J$. So Eq.(\ref{eq:A3}) reduces to
\begin{align}
  \label{eq:A6}
  f = \frac{\pi v}{\beta \mathcal J} \frac{1}{\cos \left(\pi v \left(\frac{\tau}{\beta} - \frac{1}{2}\right)\right)}
\end{align}
for $\tau > 0$. For $\tau<0$, the boundary condition at $\tau=\beta$, namely, Eq.(\ref{eq:A5}), is replaced by the boundary condition at $\tau = -\beta$, which reads
\begin{align}
  \label{eq:A7}
  \frac{\pi v}{\beta \mathcal J} \frac{1}{\cos (\pi v (1+\phi_0))} = 1.
\end{align}
Combining this and Eq.(\ref{eq:A4}), we obtain $\phi_0 = -1/2$. Substituting this into Eq.(\ref{eq:A4}), again, we obtain Eq.(\ref{eq:0.2}) and reduce Eq.(\ref{eq:A3}) to
\begin{align}
  \label{eq:A8}
  f = \frac{\pi v}{\beta \mathcal J} \frac{1}{\cos\left( \pi v \left(-\frac{\tau}{\beta} - \frac{1}{2}\right)\right)}.
\end{align}
Equations (\ref{eq:A6}) and (\ref{eq:A8}) give Eq.(\ref{eq:19}).

\end{document}